\documentclass[acmtog,authorversion]{acmart}
\usepackage{microtype}

\usepackage{booktabs} 
\usepackage{bm}
\usepackage{overpic}
\usepackage{siunitx}
\usepackage{wrapfig}
\usepackage[utf8]{inputenc}
\usepackage{multirow}
\usepackage{colortbl}
\usepackage{xr}
\newlength\savedwidth
\newcommand{\whline}[1]{\noalign{\global\savedwidth\arrayrulewidth
                              \global\arrayrulewidth #1} %
                     \hline
                     \noalign{\global\arrayrulewidth\savedwidth}}

\definecolor{liquidColor}{RGB}{131, 139, 197}
\definecolor{solidColor}{RGB}{231, 65, 93}
\definecolor{nodeColor}{RGB}{175, 167, 14}

\setcitestyle{square}

\usepackage[ruled]{algorithm2e} 

\renewcommand\footnotetextcopyrightpermission[1]{} 
\definecolor{figred}{rgb}{1,0,0}
\definecolor{figgreen}{rgb}{0,0.6,0}
\definecolor{figblue}{rgb}{0,0,1}
\definecolor{figpink}{rgb}{1,0.63,0.63}

\newcommand{\todo}[1]{}

\begin{document}

\title{Addressing Troubles with Double Bubbles: Convergence and Stability at Multi-Bubble Junctions}
\author{Yun (Raymond) Fei}
\orcid{0000-0001-8553-1377}
\affiliation{%
  \institution{Columbia University}
  \department{Computer Science}
  \city{New York}
  \state{NY}
  \postcode{10027}
  \country{USA}}
\author{Christopher Batty}
\affiliation{%
  \institution{University of Waterloo}
  \department{Computer Science}
  \city{Waterloo}
  \state{ON}
  \postcode{N2L 3G1}
  \country{Canada}}
\author{Eitan Grinspun}
\affiliation{%
  \institution{Columbia University}
  \department{Computer Science}
  \city{New York}
  \state{NY}
  \postcode{10027}
  \country{USA}}

\renewcommand\shortauthors{Fei, Y. et al}

\maketitle
\begin{figure*}[t]
 \centering
\includegraphics[width=1.0\textwidth]{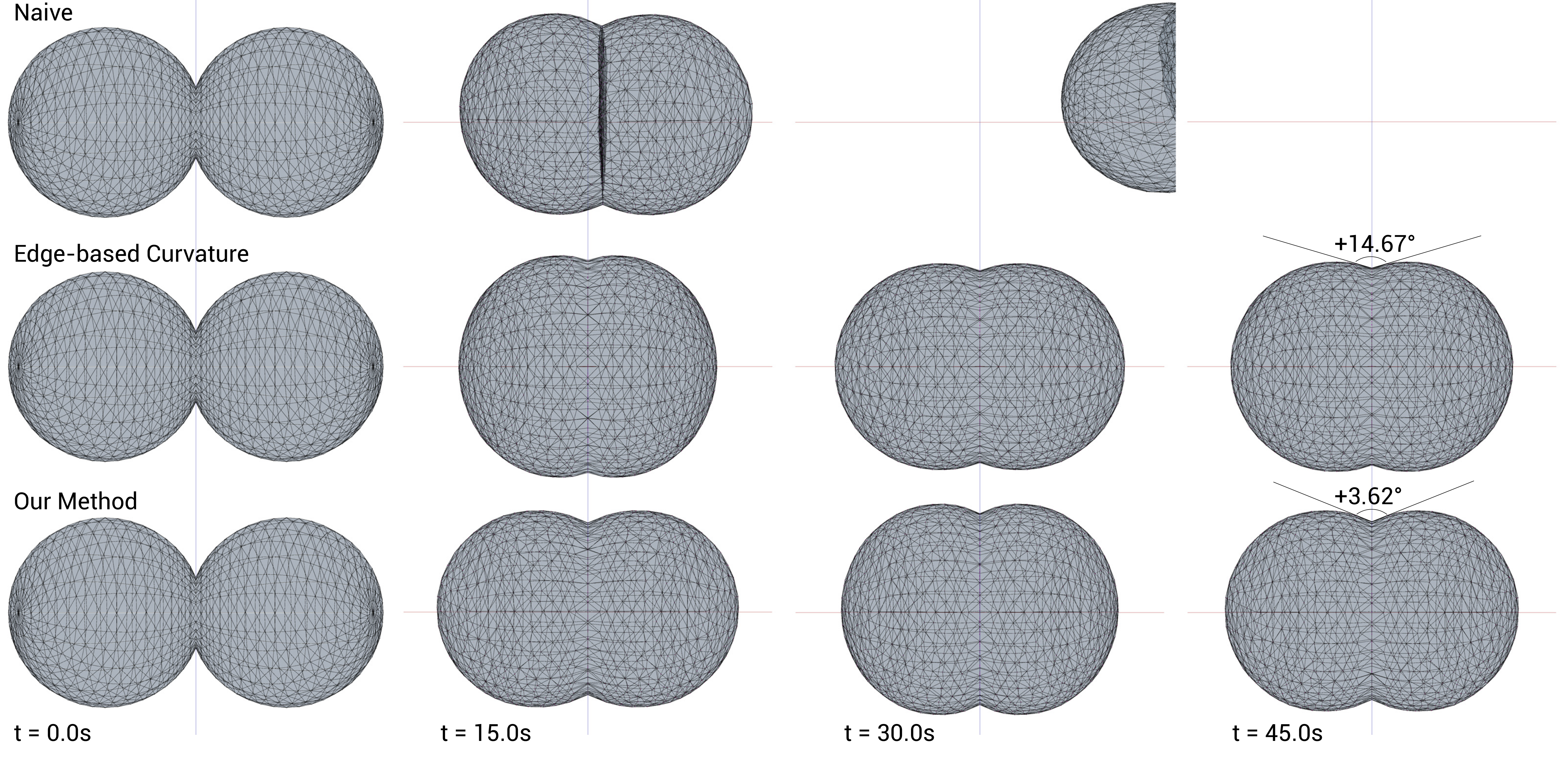}
 \caption{
\textbf{Top}: vertex area is separately calculated inside each region, without any special treatment at the junction -- the unstable double bubble drifts out of frame; 
\textbf{Middle}: vertex area at the junction is calculated using edge length times a constant -- the rest angle deviates significantly from $120^\circ$; 
\textbf{Bottom}: vertex area computed using our method where vertex area is averaged from each incident domain -- the angle deviation is much closer to zero (and converges under refinement).
\label{fig:combined}
}
\end{figure*}

\section{Introduction}
In this report we discuss and propose a correction to a convergence and stability issue occurring in the work of Da et al.~\shortcite{da2015double}, in which they proposed a numerical model to simulate soap bubbles.

In the original implementation of their work, convergence of the geometry towards equilibrium surfaces did not behave as expected. Soap foam should converge to a configuration described by Plateau's laws~\shortcite{plateau1873statique}, where the dihedral angles of faces incident to Plateau borders (triple junction edges) should be 120 degrees. However, the existing method of Da et al.\ converges to a steady state that fails to satisfy this condition: the angles differ noticeably from 120 degrees. For example, in a simple double-bubble test case, the dihedral angles of the pair of faces incident to the outer air region are approximately 135 degree, as seen in Figure~\ref{fig:combined}-middle.

We traced this issue to a flaw in the original implementation, which applied a special treatment in computing the appropriate vertex area at the triple-junction that maintains stability of surface tension forces. Unfortunately, this treatment also sacrificed the correct convergence behavior of soap foam. This undesired behavior was first observed by Ishida et al.~\shortcite{ishida2017hyperbolic}. In the following we briefly describe the problem and introduce a new method to compute the vertex area, which simultaneously maintains stability and yields the correct Plateau border angles.

\section{Integral and pointwise mean curvatures}
\begin{figure}[t]
 \centering
\includegraphics[width=1.0\columnwidth]{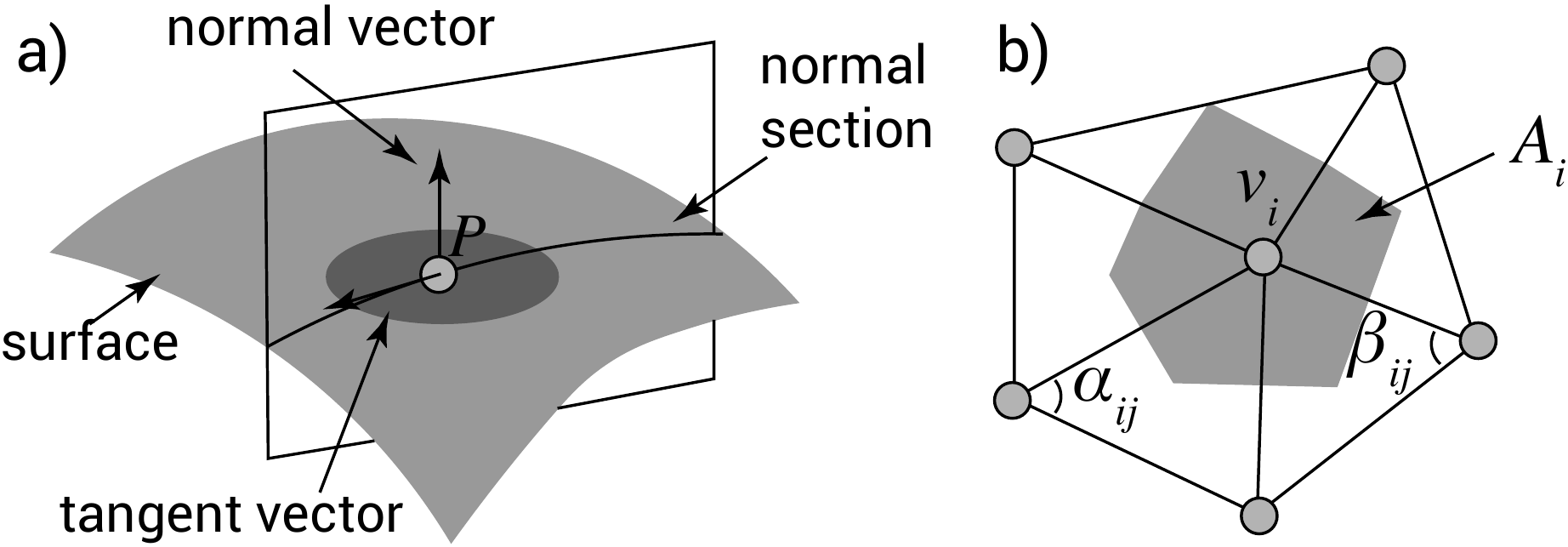}
 \caption{
\textbf{(a)}: A surface is intersected with a plane containing the normal vector. In the continuous setting, we can compute the mean curvature of a point on this surface by integrating through all the angles (dark gray circle); 
\textbf{(b)}: In the discrete setting, we can use the cotan-formula to compute the mean curvature from all nearby triangles, normalizing by the vertex (Voronoi) area. 
\label{fig:geo}
}
\end{figure}
To understand the problem, we begin with a discussion of mean curvature. In the continuous setting (Figure~\ref{fig:geo}a), the mean curvature at a point $P$ can be interpreted as an integral of the signed curvature on the normal section curve over all directions (represented by the dark gray circle in Figure~\ref{fig:geo}a). 

In the discrete setting of a numerical simulation, only finitely many triangles are used to approximate the surface. The pointwise discrete mean curvature of the surface can be computed for each (volumetric) region with the following formula~\cite{sullivan2008curvatures} for vertex $v_i$,
$$
H(v_i)=\frac{1}{2A_i}\left\|\sum_j\kappa_j\right\|
$$
where $\kappa_j$ is the edge curvature of the $j$-th edge connected to vertex $v_i$, and $H(v_i)$ is normalized by the vertex area $A_i$ (Figure~\ref{fig:geo}b). The edge curvature is given by $|e|\theta$~\cite{cohen2003restricted}. The normalization by area converts integrated curvature into the pointwise curvature which is then applied as a force by Da et al. 

For two volumetric regions sharing an interface the \emph{absolute} pointwise mean curvatures are trivially the same; the two sides of the surface are geometric complements and therefore differ only in sign, leading to a natural force balance at equilibrium. However, when three or more regions are present, a vertex may lie on a triple (or higher-order) junction, and computing discrete mean curvature separately in each region at the junction becomes problematic. More specifically, since the discrete mean curvature is proportional to the inverse of the vertex area, the choice of how to define the vertex area per region has a critical effect, as we describe below.

\section{A problematic case}
\begin{figure}[t]
 \centering
\includegraphics[width=1.0\columnwidth]{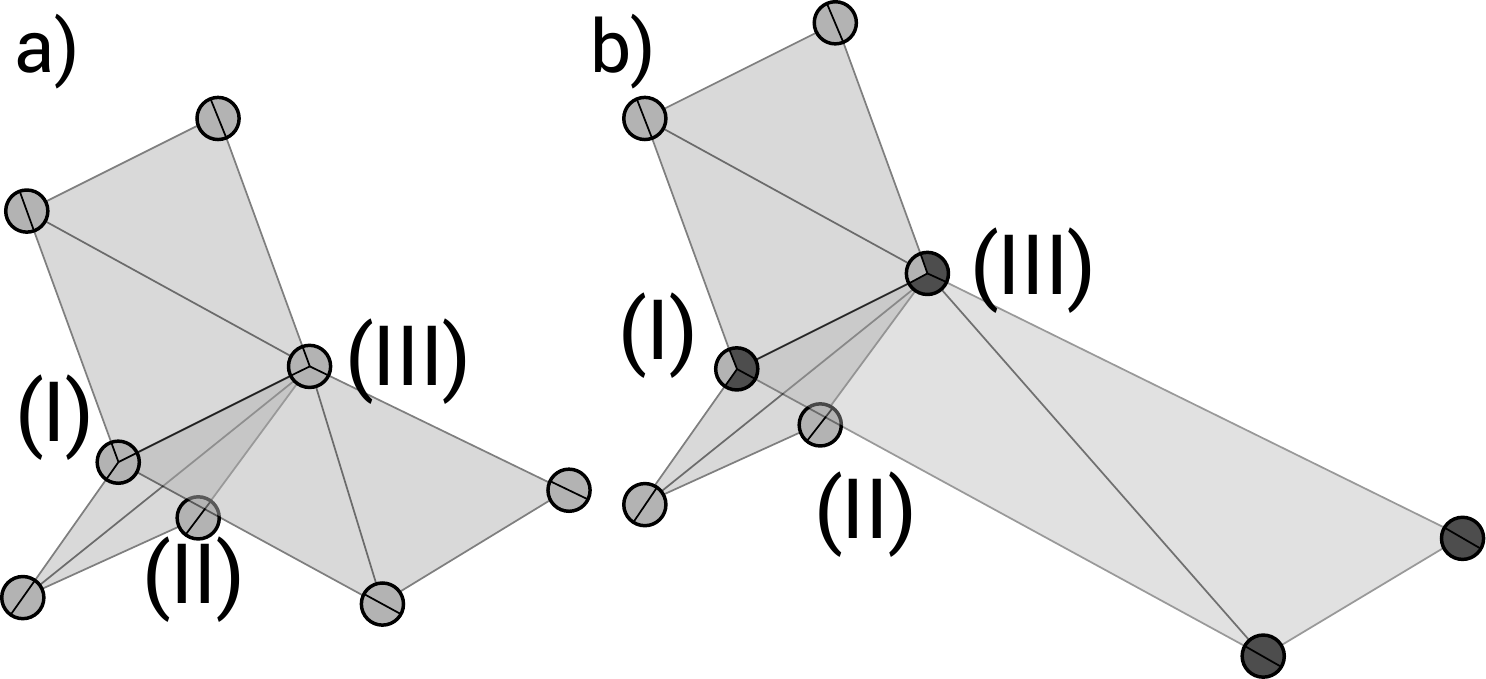}
 \caption{
\textbf{(a)}: A triple junction comprised of triangles. On each vertex the mean curvatures per region are computed, with the subdivided small disks indicating the set of regions involved. 
\textbf{(b)}: The same triple junction but with the triangles between region (II) and (III) stretched. In this case, under the na\"{i}ve discretization (only) the discrete mean curvatures in region (II) and (III) of the vertices on the stretched triangles are affected (marked with dark gray), yet the triple junction geometry has not changed.
\label{fig:stretch}
}
\end{figure}
We will consider the triple-junction in Figure~\ref{fig:stretch} as our motivating example. If one computes the mean curvatures at the junction separately per volumetric region using only the triangles comprising each region's surface, then the area of a triangle on one particular branch of the junction contributes to the curvature estimate for only the two volumes that share that triangle (Figure~\ref{fig:stretch}b). The curvature estimate for the third region is entirely independent of that triangle's area; this implies that if the mesh is initially in a numerical equilibrium and that triangle is then stretched (or remeshed) in a manner that doesn't change the angles at the triple-junction, the observed curvature values of the two incident volumes will change accordingly, but that of the third will not. That is, force balance will be lost, despite no change in the actual angles at the junction! In the limit of mesh refinement the vertex area conceptually shrinks to zero ensuring a valid discretization, but in any discrete simulation the delicate force balance is often quickly destroyed. This problem compounds from one timestep to the next as the resulting forces attempt to compensate for changing incident triangle areas by adapting the junction angles, leading to further instability.  As a consequence, when simulating double bubbles, the bubbles will incorrectly accelerate and drift away (Figure~\ref{fig:combined}-top).

\section{A preliminary solution}
In the original implementation of Da et al.~\shortcite{da2015double}, the authors noticed this problem and adopted a simple solution that modifies the vertex area computation at the triple junction: the vertex area used for all the incident regions is set to be the shared edge length along the triple junction times a global constant, which is set to be a fixed fraction of the mean edge length over the entire mesh ($0.5$ is used in the original implementation). In this way, the discrete mean curvatures computed at the triple junction reflect the relevant angles, but are independent of the areas of the particular triangulation.

This simple solution can indeed stabilize the simulation since the mean curvatures will not be affected by otherwise irrelevant disturbances in the local triangulation; for example, in-plane vertex-smoothing of nearby points no longer suddenly upsets the balance. Nevertheless, a global constant times the edge length cannot well approximate the actual local vertex area. In the particular example of our double-bubble, this ``effective vertex area" for the exterior region at the triple-junction is larger than the actual value computed directly from the triangles, while the vertex area computed inside the bubble is smaller. As a result, the mean curvature outside the bubble is smaller and yields a larger surface tension force, which causes the dihedral angle between the bubbles to be larger than the desired equilibrium value (Figure~\ref{fig:combined}-middle).

\section{Our solution}
\begin{table}[h]
\centering
\caption{\textbf{Angular deviation versus resolution}: the observed angular deviation from $120^\circ$ at the triple junction decreases as the resolution of the mesh increases, confirming that our method converges under refinement.}
\label{tab:errorres}
\begin{tabular}{ccc}
\whline{1.0pt}
\# Subdivision & \# Faces  & \multirow{2}{*}{Angular Deviation} \\ 
(per bubble)   & (initial) &      \\ \hline
\whline{0.8pt}
16             & 880       & $12.79^{\circ}$                 \\ \rowcolor[gray]{0.93}
24             & 1,992     & $9.35^{\circ}$                  \\ 
48             & 8,016     & $4.15^{\circ}$                  \\ \rowcolor[gray]{0.93}
96             & 32,160    & $2.65^{\circ}$                  \\ 
\whline{1.0pt}
\end{tabular}
\end{table}
Although the remeshing process can cause unpredictable changes in the local triangulation, we identified a strategy that can always be safely applied: we evaluate the vertex areas per region, determine their average, and use that value for normalization when computing each of the per-region discrete curvatures. This technique is simple to implement, incurs minimal additional cost, and can be justified as follows.

Under this strategy, since the normalizing vertex areas used in computing the curvature for each region at the vertex are the same, when a nearby vertex in any region is disturbed, the change is immediately reflected in the computed mean curvatures for all the incident regions. This greatly improves stability, similar to the fix by Da et al. However, unlike the correction used by Da et al., the vertex areas used in our method faithfully reflect the actual areas of the surrounding triangles, rather than an arbitrary global constant. Hence convergence can be assured since valid mean curvatures are computed for the vertex in each region. In other words, the normalization of curvatures at the triple junction is now scaled appropriately by local area, consistent with how the discrete curvatures are scaled by area on other (manifold, non-junction) parts of the mesh.

In Figure 1 bottom, we demonstrate that our method produces a stable simulation of the double-bubble, and that bubbles at equilibrium have a dihedral angle much closer to the theoretical value of $120^{\circ}$. In Table~\ref{tab:errorres}, we demonstrate that our discretization converges: as the resolution of the mesh increases, the angular deviation decreased dramatically from $12.79^{\circ}$ to $2.65^{\circ}$, indicating that the dihedral angles in our simulation converges towards the theoretical value. The angular deviation was computed as the root-mean-square deviation compared with $120^{\circ}$, across all the dihedral angles at the triple junction (consistent with Ishida et al.~\shortcite{ishida2017hyperbolic}).

\bibliographystyle{ACM-Reference-Format}
{
\setlength{\topsep}{0ex}
\setlength{\itemsep}{0ex}
\setlength{\partopsep}{0ex}
\setlength{\parsep}{0ex}
\setlength{\parskip}{1ex}
\bibliography{ref}
}

\end{document}